\documentclass[
    ,final            
  ]
  {aipproc}

\layoutstyle{8x11double}

\usepackage{amsmath}
\usepackage{amsfonts}
\usepackage{amssymb}
\usepackage{graphicx}
\usepackage[english]{babel}
\usepackage{wasysym}
\usepackage{units}

\newcommand{\reff}[1]{(\ref{#1})}

\begin{document}

\title{Scalar-tensor analysis of an exponential Lagrangian for the Gravitational Field}


\author{O.M. Lecian}{
address={ICRA -- International Center for Relativistic Astrophysics\\
Department of Physics - Università di Roma ``Sapienza'', Piazza A. Moro, 
5 (00185), Rome, Italy\vspace{0.2cm}},
}

\author{G. Montani}{
altaddress={ICRA -- International Center for Relativistic Astrophysics\\
Department of Physics - Università di Roma ``Sapienza'', Piazza A. Moro, 
5 (00185), Rome, Italy\vspace{0.2cm}},
address={ENEA -- C.R. Frascati (Department F.P.N.), Via Enrico Fermi, 45 
(00044), Frascati (Rome), Italy\\ICRANet -- C. C. Pescara, Piazzale 
della Repubblica, 10 (65100), Pescara, Italy\vspace{0.2cm}\\{\footnotesize\ttfamily lecian@icra.it\qquad montani@icra.it}
}
}

\begin{abstract}
Within the scheme of modified gravity, an exponential Lagrangian density will be considered, and the corresponding scalar-tensor description will be addressed for both positive and negative values of the cosmological constant.\\ 
For negative values of the cosmological term, the potential of the scalar field exhibits a minimum, around which scalar-field equations can be linearized. The study of the deSitter regime shows that a comparison with the modified-gravity description is possible in an off-shell region, i.e., in a region where the classical equivalence between the two formulations is not fulfilled. Furthermore, despite the negative cosmological constant, an accelerating deSitter phase is predicted in the region where the series expansion of the exponential term does not hold.\\
For positive values of the cosmological constant, the quantum regime is analyzed within the framework of Loop Quantum Cosmology.\end{abstract}

\maketitle

\section{Introduction}
One of the most challenging tasks in the modern understanding of the Universe evolution is the explanation of the present value of the cosmological constant. The accelerating character of the present Universe, supported by observational data on the recession of SNIA , can be described by a negative-pressure contribution, and the analysis of CMB suggests that the so-called Dark Energy has reliably the features of a cosmological constant, which corresponds to about 70 percent of the critical density of the Universe. Such an amount of the cosmological term is relevant for the actual dynamics, but extremely smaller than the vacuum value. Estimations of the vacuum energy yield indeed the Planckian value, corresponding to $10^{120}$ times the observed numbers. This striking contradiction between the theoretical predictions and the actual value suggests that, if the Universe acceleration is really due to a cosmological constant, then a precise mechanism of cancellation must be fixed for the vacuum energy density. Since no fundamental theory provides a convincing explanation for such a cancellation, it is naturally expected to find it from specific features of the field dynamics. The main proposals for such a behavior can be classified in two groups: those that make explicitly presence of matter, and those that relay on modifications of the Friedmann dynamics.\\ 
Here, we address a mixture of these two points of view, with the aim of clarifying how the ''non-gravitational'' vacuum energy affects so weakly the present Universe dynamics (see \cite{lm07} and the references therein). Indeed, our model is not aimed at showing that the present Universe acceleration is a consequence of non-Einsteinian dynamics of the gravitational field, but at outlining how it can be recognized from a vacuum-energy cancellation. Such a cancellation must take place in order to deal with an expandable Lagrangian term and must concern the vacuum-energy density as far as we build up the geometrical action only by means of fundamental units, i.e., the cancellation that takes place between the intrinsic term and the effective vacuum energy leaves a relic term, of order $10^{-120}$ times the present Universe Dark Energy, much smaller than the original. If we want to build up a generalized gravitational action, which depends on the Planck length as the only parameter, then the geometrical components of an exponential one contain a cosmological term too, whose existence can be recognized as soon as we expand the exponential form in Taylor series of its argument. For Planckian values of the fundamental parameter of the theory (requested by the cancellation of the vacuum-energy density), as far as the Universe leaves the Planckian era and its curvature has a characteristic length much greater than the Planckian one, the series expansion of the corresponding Lagrangian density holds, and reproduces General Relativity (GR) to a high degree of approximation, so that most of the thermal history of the Universe is unaffected by the generalized theory, but for the fact that the deSitter solution exists in presence of matter only for a negative ratio between the vacuum-energy density and the intrinsic cosmological term, $\epsilon_{vac}/\epsilon_{\Lambda}$. This fact looks like a fine-tuning, especially if we take, as we will do below, a Planckian cosmological constant. The vacuum-energy density is expected to be smaller than the Planckian one by a factor $\mathcal{O}(1)\times\alpha^{4}$, where $\alpha<1$ is a parameter appearing in non-commutative models.\\ 
The paper is organized as follows.\\
After introducing and discussing the formalisms of modified gravity and scalar-tensor models, we briefly review the main theoretical aspects of the vacuum-energy problem. We then analyze the exponential action in the Jordan frame, where the Einsteinian regime can be recovered after a series expansion. Nevertheless, two different possibilities are found: the series expansion either does not hold or brings puzzling predictions about the cosmological term. Correspondingly, in the first case, an unlikely implication would appear when dealing with a non-Einsteinian physics on all astrophysical scales, and, in the second case, the expansion is only possible in the region $\Lambda>>R$, i.e., in the region where the cosmological constant dominates the dynamics, but for the fact that $R$ should be the same order of $\Lambda$. This contradiction can only be solved if a suitable cancellation mechanisms is hypothesized: here we find the constraint on the ratio $\epsilon_{vac}/\epsilon_{\Lambda}\ll1$ in the deSitter regime in presence of matter. The analysis of the corresponding scalar-tensor model helps us shed light on the physical meaning of the sign of the cosmological term. Two possibilities are taken into account. For positive values of the cosmological constant, we establish to what extent we can compare our model with the theoretical framework of Loop Quantum Cosmology, in which the cosmological singularity is removed. For negative values of the cosmological constant, the potential of the scalar field exhibits a minimum, around which field equations can be linearized. The study of the deSitter regime shows that a comparison with the modified-gravity description is possible in an off-shell region, i.e., in a region where the classical equivalence between the two formulations is not fulfilled.\\
A proposal for the solution of the puzzle is eventually exposed in Section 6, where the Universe acceleration is related to the vacuum energy through the introduction of the dimensionless parameter $\delta$, which acts like a compensating factor between the energy density associated to the cosmological constant and that estimated for the vacuum energy in presence of a cut-off.
Concluding remarks follow.
\section{Modified and scalar-tensor gravity}
The action describing the gravitational field coupled to matter reads
\begin{equation}
S=S_{EH}+S_{M}\equiv -\frac{c^{3}}{16\pi G}\int d^{4}x \sqrt{-g} R +\frac{1}{c}\sum _{f}\int d^{4}x\sqrt{-g} L_{f},
\end{equation}
i.e., the Einstein-Hilbert (EH) action associated to the metric tensor $g_{\mu\nu}$ and that of all matter fields $f$. After variation with respect to $g_{\mu\nu}$, the well-known Einstein equations follow
\begin{equation}
R_{\mu\nu}-\frac{1}{2}g_{\mu\nu}R=\frac{8\pi G}{c^{4}}T_{\mu\nu},
\end{equation}
where $T_{\mu\nu}$ is the stress-energy tensor.\\
Replacing the Ricci scalar $R$ with a generic function $f(R)$,
\begin{equation}\label{fdr}
S_{EH}\rightarrow -\frac{c^{3}}{16\pi G}\int d^{4}x \sqrt{-g} f(R),
\end{equation}
allows one to obtain the generalized Einstein equations\footnote{The $00$ and $ii$ components of \reff{generale} can be obtained also considering the FRW line element 
\begin{equation}\label{line}
ds^{2}=N(t)^{2}dt^{2}-a(t)^{2}dl^{2},
\end{equation}
where $N(t)$ is the lapse function, and $a(t)$ the cosmic scale factor: under the finite-volume assumption, the total action rewrites
\begin{equation}
S= -\frac{Vc^{4}}{16G\pi}\int dt Na^{3} f(R)-V\int dt Na^{3}\epsilon (t).
\end{equation}
Variation with respect to $N$ leads to the standard Euler-Lagrange equation $\frac{\partial L}{\partial N}-\frac{d}{dt}\frac{\partial L}{\partial \dot{N}}=0$, which, in the synchronous reference frame, i.e., $N=1$, reads
\begin{equation}\label{one}
\frac{1}{2}f+3f'\frac{\ddot{a}}{a}-3f''\frac{dR}{dt}\frac{\dot{a}}{a}=-\frac{8G\pi}{c^{4}}\epsilon ,
\end{equation}
and is the same as the $00$-component of the generalized Einstein equations for the FRW metric.\\
Variation with respect to $a$ leads to the generalized Euler-Lagrange equation $\frac{\partial L}{\partial a}-\frac{d}{dt}\frac{\partial L}{\partial \dot{a}}+\frac{d^{2}}{dt^{2}}\frac{\partial L}{\partial \ddot{a}}=0$, and, for $N=1$, we find
\begin{align}\label{two}
&-\frac{1}{2}f+f'\left[-\frac{\ddot{a}}{a}-2\frac{\dot{a}^{2}}{a^{2}}-2\frac{k}{a^{2}}\right]+2f''\frac{dR}{dt}\frac{\dot{a}}{a}+\nonumber \\
&+f'''\left(\frac{dR}{dt}\right)^{2}+f''\frac{d^{2}R}{dt^{2}}=-\frac{8G\pi}{c^{4}}p.
\end{align}
Combining together \reff{one} and \reff{two}, the equation for the Universe acceleration cab be found.} in the so-called Jordan frame
\begin{align}\label{generale}
&-\frac{1}{2}g_{\mu\nu}f(R)+f'(R)R_{\mu\nu}-\nabla_{\nu}\nabla_{\nu}f'(R)+\nonumber\\
&+g_{\mu\nu}\nabla_{\rho}\nabla^{\rho}f'(R)=\frac{8\pi G}{c^{4}}T_{\mu\nu},
\end{align}
where $f'(R)\equiv df(R)/dR$. This way, the model is still invariant under 4-diffeomorphisms, but higher-order contributions could become relevant for high values of space-time curvature.\\
It can be demonstrated that the non-linear Lagrangian \reff{fdr} can be cast in a dynamically-equivalent form, i.e., that of a scalar field in GR (with a rescaled metric), by introducing two Lagrange multipliers, and then performing a suitable conformal transformation \cite{noji03}. The two Lagrange multipliers $A$ and $B$ allow one to rewrite (\ref{fdr}) as
\begin{equation}\label{mult}
S=\frac{1}{k^{2}}\int d^{4}x\sqrt{-g}\left[B(R-A)+f(A)\right],
\end{equation}
where variation with respect to $B$ leads to $R=A$, while variation with respect to $A$ gives the identity $B=f'(A)$, or, equivalently, $A=g(B)$. It is possible to eliminate either $A$ or $B$ from (\ref{mult}), thus obtaining
\begin{equation}\label{subst}
S=\frac{1}{k^{2}}\int d^{4}x\sqrt{-g}\left[B(R-g(B))+f(g(B))\right]
\end{equation}
or
\begin{equation}\label{elimination}
S=\frac{1}{k^{2}}\int d^{4}x\sqrt{-g}\left[f'(A)(R-A)+f(A)\right],
\end{equation}
respectively. Equations (\ref{subst}) or (\ref{elimination}) are equivalent, at least from a classical point of view, and are usually referred to as the Jordan-frame action in presence of the two auxiliary fields. Furthermore, the conformal scaling of the metric tensor  
\begin{equation}\label{conf}
g_{\mu\nu}\rightarrow e^{\phi}g_{\mu\nu}
\end{equation}
allows one to cast the previous results in the Einstein frame. For the particular choice $\phi=-\ln f'(A)$, action (\ref{elimination}) reads
\begin{equation}\label{scaltens}
S=\frac{1}{k^{2}}\int d^{4}x\sqrt{-g}\left[R-\frac{3}{2}g^{\rho\sigma}\partial_{\rho}\phi\partial_{\sigma}\phi-V(\phi)\right],
\end{equation}
where $V(\phi)=A/f'(A)-f(A)/f'(A)^{2}$, i.e., it describes a scalar field minimally-coupled to the rescaled metric\footnote{If an external matter fluid is taken into account, the pertinent stress-energy tensor $T_{\mu\nu}$ associated to the energy density $\epsilon$, the pressure $p$ and the four-velocity $u_{\mu}$, $T_{\mu\nu}=(\epsilon+p)u_{\mu}u_{\nu}-pg_{\mu\nu}$, has to be rescaled as 
\begin{equation}\label{conf2}
T_{\mu\nu}\rightarrow e^{-\phi}\hat{T}_{\mu\nu}, \quad T^{\mu}_{\quad\nu}\rightarrow e^{-2\phi}\hat{T}^{\mu}_{\quad\nu},
\end{equation}
according to the conformal transformations induced by (\ref{conf}), i.e.,
\begin{equation}\label{conf3}
u_{\mu}\rightarrow e^{\phi/2}\hat{u}_{\mu}, \quad \epsilon\rightarrow e^{-2\phi}\hat{\epsilon}, \quad p\rightarrow e^{-2\phi}\hat{p}.
\end{equation}}.\\
Moreover, once in the Einstein frame, a back conformal transformation can be performed, in order to obtain a vanishing kinetic term for the scalar field, in the so-called fluid description. This way, the scalar field becomes an auxiliary field, which can be eliminated by means of the equation of motion\footnote{While the mathematical meaning of Lagrange multipliers and conformal transformations, which establishes the mathematical equivalence between the models, is well-understood, the physical interpretation of these manipulations needs further clarification. In fact, besides the transformation that maps the Jordan frame into the Einstein frame, there exist infinitely many conformally-related models. In the Jordan frame, gravity is entirely described by the metric tensor, but, in the Einstein frame, the scalar field minimally coupled to gravity represents and additional degree of freedom, which corresponds to the higher-order corrections, and could be interpreted as a non-metric contribution to the EH action \cite{magn}. Additionally, the physical equivalence between the models cannot be discussed a priori: throughout this paper we will try to investigate the role and the properties of the matter field in the determination of cosmological solutions.}.
\section{The vacuum-energy problem}
As well known, the vacuum-energy density associated to a massless quantum field is a diverging quantity unless an appropriate normal ordering (which, on curved space-time, would depend on the metric properties of the manifold) can be found; however, if we fix a cut-off on the momentum variable, $P_{max}=\alpha \frac{\hbar}{l_{pl}}$ ($\alpha$ being a dimensionless parameter of order unity) then the vacuum energy density can be estimated as follows
 \begin{equation}\label{ved}
\epsilon_{vac}=\int_{0}^{P_{max}} \frac{d^{3}p}{\hbar^{3}}cp=\int_{0}^{P_{max}} \frac{4\pi p^{2}dp}{\hbar^{3}}cp=\pi\alpha^{4}\epsilon_{pl},
 \end{equation}
where $\epsilon_{pl}\equiv\hbar c/l_{pl}^{4}$, i.e., we deal with a vacuum-energy density of a Planck-mass particle per Planck volume. A more rigorous understanding for the parameter $\alpha$ comes out from an approach based on GUP. Such theories implement modified canonical operators obeying the generalized relation\footnote{This commutation relation can be recognized on the ground of fundamental properties of the Minkowski space in presence of a cut-off, but it also comes out from quantum-gravity and string-theory approaches. As a consequence of non-commutative models, we deal with a notion of minimal length associated to a particle state. For instance, in the case of a non-relativistic particle, we get the following limit for its wave-length
\begin{equation}
\lim_{E\rightarrow\infty}\lambda(E)=\frac{4}{\alpha} l_{pl},
\end{equation}
$E$ being the energy of the particle.\\
For a discussion of a maximum value for a relativistic-particle momentum at Planck scales, in the context of the k-Poincar\'e algebra, see \cite{sdfg}. But it is worth noting that, in our case, the discussion above must be referred to a flat FRW background, and, therefore, all the observables correspond with physical quantities corrected by the presence of the scale factor.\\}
\begin{equation}
[x,p]=i\hbar(1+\frac{1}{\alpha^{2}}\frac{G}{c^{3}\hbar}p^{2}).
\end{equation}
However, no evidence appears today for such a huge cosmological term. Nonetheless, when estimating this observed cosmological term, it is immediate to recognize that it is extremely smaller than the cut-off value. In fact, for the observed value of the constant energy density, we get the estimation
\begin{equation}\label{poi}
\epsilon_{today}\sim 0.7\epsilon_{0}\cong \frac{2c^{2}H_{0}^{2}}{8\pi G}=\frac{1}{4\pi\alpha^{4}}\left(\frac{l_{pl}}{L}\right)^{2},
\end{equation}
where $\epsilon_{0}$ denotes the present Universe critical density $\epsilon_{0}\sim\mathcal{O}(10^{-29})g cm^{-3}$, $H_{0}\sim 70 Km s^{-1} Mpc^{-1}$ the Hubble constant, and $\epsilon_{today}$ the present value of the vacuum-energy density; since $L_{H}\equiv cH_{0}\sim\mathcal{O}(10^{27}cm^{-1})$, we see that a large factor $10^{-120}$ appears in (\ref{poi}), i.e. $\epsilon_{today}\sim\mathcal{O}(10^{-120})\epsilon_{vac}$. It is well known that this striking discrepancy between the expected and the observed value of the vacuum energy constitutes one of the greatest puzzle of modern cosmology.
\section{Exponential modified gravity}
While recent observations on Supernovae seem to indicate that the Universe is now accelerating with a non-definite equation of state, data from CMB anisotropy suggest that the most appropriate characterization of such an equation of state would be that of a cosmological term, i.e., $p\sim-\epsilon$. This way, the appearance of a cosmological constant, which is usually added by hand to the EH action, could originate from the series expansion of an $f(R)$ scheme. Nevertheless, one would have to fix an infinite number of coefficients in order to deal with such an expansion. This problem can be overcome by the assumption that only one characteristic length, provided by observational data, should fix the dynamics, i.e., the cosmological constant $\Lambda$, apart form the Planck length $l_{pl}\equiv \sqrt{(G\hbar/c^{3})}$. The most natural choice of the generalized model is the following:  
\begin{equation}\label{cat5}
f(R)=\lambda e^{\mu R},
\end{equation}
where $\lambda$ and $\mu$ are two free parameters available for the theory. Nonetheless, the comparison of the two lowest orders of the series expansion with the EH action plus a cosmological term, $L=-\frac{\hbar}{16 \pi l^{2}_{P}}\left( R+2\Lambda \right)$, leads to the straightforward identifications
\begin{equation}
\lambda=2\Lambda,\ \ \mu=\frac{1}{2\Lambda}.
\end{equation}
The analysis of the deSitter regime helps us gain insight into this generalized FRW dynamics. To this aim, we consider the cosmic scale factor $a$ and the dimensionless parameter $x$, such that  
\begin{equation}
a=a_{0} e^{\sigma t},\ \ a_{0}=const,\ \ \sigma =const., \ \ x\equiv \frac{6\sigma^2}{\Lambda c^2}:
\end{equation}
in this case, the Ricci scalar reads becomes $R=-12\sigma^{2}/c^2$.\\
If external matter is absent, the value $x=-2$ implies that the series expansion of the exponential function does not hold, and we are dealing with the full non-perturbative regime with respect to Einstein gravity.\\
The series expansion can be performed if an external matter field is added. The introduction of this (rather unphysical) matter field will be illustrated to be an eligible candidate for the explanation of the mechanism that induces as the (quasi) cancellation of the Plank-scale vacuum-energy density. In fact, if an external matter field is introduced, the Friedmann equation rewrites 
\begin{equation}\label{sopra}
\epsilon=-\epsilon_{\Lambda}e^{-x}\left(1+\frac{x}{2}\right),\ \ \epsilon_{\Lambda}\equiv\frac{c^{4}\Lambda}{8\pi G},
\end{equation} 
where the energy density would acquire a negative sign. Even though this tantalizing negative sign could be removed by expanding the exponential term for small values of $x$, thus obtaining the usual Friedmann equation for matter and geometry, i.e.,
\begin{equation}\label{qwert}
\epsilon=\epsilon_{\Lambda}\left( \frac{x}{2}-1\right)\Rightarrow \sigma^{2}=\frac{8\pi G}{3c^{2}}\left( \epsilon+\epsilon_{\Lambda}\right),
\end{equation}
the inconsistency shows up again when \reff{qwert} is restated as 
\begin{equation}\label{14}
x=2\left(\frac{\epsilon}{\epsilon_{\Lambda}}+1\right).
\end{equation}
In fact, eq. \reff{14} implies $x>2$, in clear contradiction with the hypothesis $x<<1$, which allows for the series expansion\footnote{This hypothesis can however be recovered if one assumes a negative ratio $\frac{\epsilon}{\epsilon_\Lambda}$, i.e., $\Lambda<0$.\\
For the choice of a negative cosmological constant, i.e., $\Lambda\equiv-\mid\Lambda\mid$, since the Friedmann equation rewrites, in vacuum, $\sigma^{2}=\frac{c^{2}\mid\Lambda\mid}{3}$, a deSitter evolution is predicted is presence of a negative cosmological constant.\\
So far, a negative cosmological constant is needed to recover a series expansion in the low-curvature limit, thus allowing for the comparison with standard gravity. In fact, in the non-perturbative regime, the parameter $\Lambda$ could not be recognized as a cosmological term. This way, the non-observability of the vacuum energy is strictly connected with the present Universe acceleration.}. 
\section{Exponential scalar-tensor gravity}
The scalar-tensor formalism is here applied to the particular choice of the exponential Lagrangian density, in order to clarify the meaning of the relations found in the previous section.\\
The conformal scaling factor here reads
\begin{equation}
f(A)=\lambda e^{\mu A}, \quad f'(A)=e^{-\phi},
\end{equation}
where $A=-\phi/\mu$, and the potential rewrites
\begin{equation}\label{pot}
V(\phi)=-2\Lambda e^{\phi}(\phi+1). 
\end{equation}
The on-shell relation between the Einstein frame and the Jordan one is recognized in the identification $A\equiv R\Rightarrow\phi\equiv-R/(2\Lambda)$.\\
Collecting all the terms together, we get the scalar-tensor action
\begin{align}\label{st3}
& S=-\frac{c^{3}}{16\pi G} \int d^{4}x\sqrt{-g}\left[R+\right.\nonumber\\
&\left.+\frac{3}{2}g^{\mu\nu}(\partial_{\mu}\phi)(\partial_{\nu}\phi)-2\Lambda e^{\phi}(\phi+1)\right].
\end{align}
To get the right dimension of a scalar field out of \reff{st3}, the transformation $\phi\rightarrow\sqrt{\frac{16\pi G}{3c^4}}\phi$ has to be considered. This way, varying eq. \reff{st3} leads to the scalar-tensor Einstein equations in presence of a matter source described by the energy-momentum tensor $\hat{T}_{\mu\nu}$; in the case of an FRW metric and a perfect fluid as external matter, we obtain
\begin{equation}\label{frieda}
\left(\frac{\dot{a}}{a}\right)^{2}=\frac{8\pi G}{3c^{2}}\left(e^{-2\sqrt{\frac{16\pi G}{3c^{4}}}\phi}\hat{\epsilon}(t)+\frac{1}{2}\dot{\phi}^{2}+\frac{1}{2}V(\phi)\right)
\end{equation}
\begin{equation}\label{frieda1}
2\frac{\ddot{a}}{a}+\left(\frac{\dot{a}}{a}\right)^{2}=-\frac{8\pi G}{c^{2}} \left(e^{-2\sqrt{\frac{16\pi G}{3c^{4}}}\phi}\hat{p}(t)+\frac{1}{2}V(\phi)\right)
\end{equation}
\begin{equation}
\ddot{\phi}+3H\dot{\phi}+c^{2}\frac{dV(\phi)}{d\phi}=0,
\end{equation}
where $H\equiv\dot{a}/a$. Equations (\ref{frieda}) and (\ref{frieda1}), i.e., the $00$ and $ii$ components of the Einstein equations, are not independent, but linked by the rescaled continuity equation.
\subsection{$\Lambda>0$}
The potential term \reff{pot} admits an absolute maximum and a slow-rolling regime\footnote{According to the potential profile, a late-time solution $\phi(t)$ can be looked for, such that $\dot{\phi}(t)\rightarrow0$ for $\phi(t)\rightarrow\infty$, and $V(\phi)\rightarrow0$ for $\phi(t)\rightarrow-\infty$.
In absence of external matter, i.e., $\epsilon(t)=0$, the Friedmann equation (\ref{frieda}) simplifies as
\begin{equation}
\left(\frac{\dot{a}}{a}\right)^{2}=\frac{8\pi G}{3c^{2}}\frac{\dot{\phi}^{2}}{2}:
\end{equation}
after standard manipulation, one finds that the time dependence of $\phi(t)$ and $a(t)$ are
\begin{equation}\label{phitronc}
\phi=\sqrt{\frac{c^{2}}{12\pi G}}\ln\left[\sqrt{\frac{12\pi G}{c^{2}}}\frac{\dot{\phi}_{0}}{a_{0}^{3}}(t-t_{0})\right],
\end{equation}
\begin{equation}\label{scalfa}
a=\frac{\dot{\phi}_{0}}{a_{0}^{2}}\sqrt{\frac{12\pi G}{c^{2}}}t^{1/3},
\end{equation}
where $\phi_0$ and $a_0$ are integration constants. As requested, at the time $t_{0}=0$, the field (\ref{phitronc}) tends to $-\infty$. So far, it is possible to verify that the potential $V(\phi)$ and its first derivative could be neglected: in fact, its contribution at early times is of order $\mathcal{O}(t^{3/2}\ln(t+1))$, which can be ignored in the presence of the leading-order terms $\mathcal{O}(t^{-2})$ due to both $(\dot{a}/a)^2$ and $\dot{\phi}^{2}$.}. In this case, the negative ratio $\frac{\epsilon}{\epsilon_\Lambda}$ can be achieved only for the unphysical condition $\epsilon<0$.
\begin{center}
\includegraphics[width=0.5\textwidth]{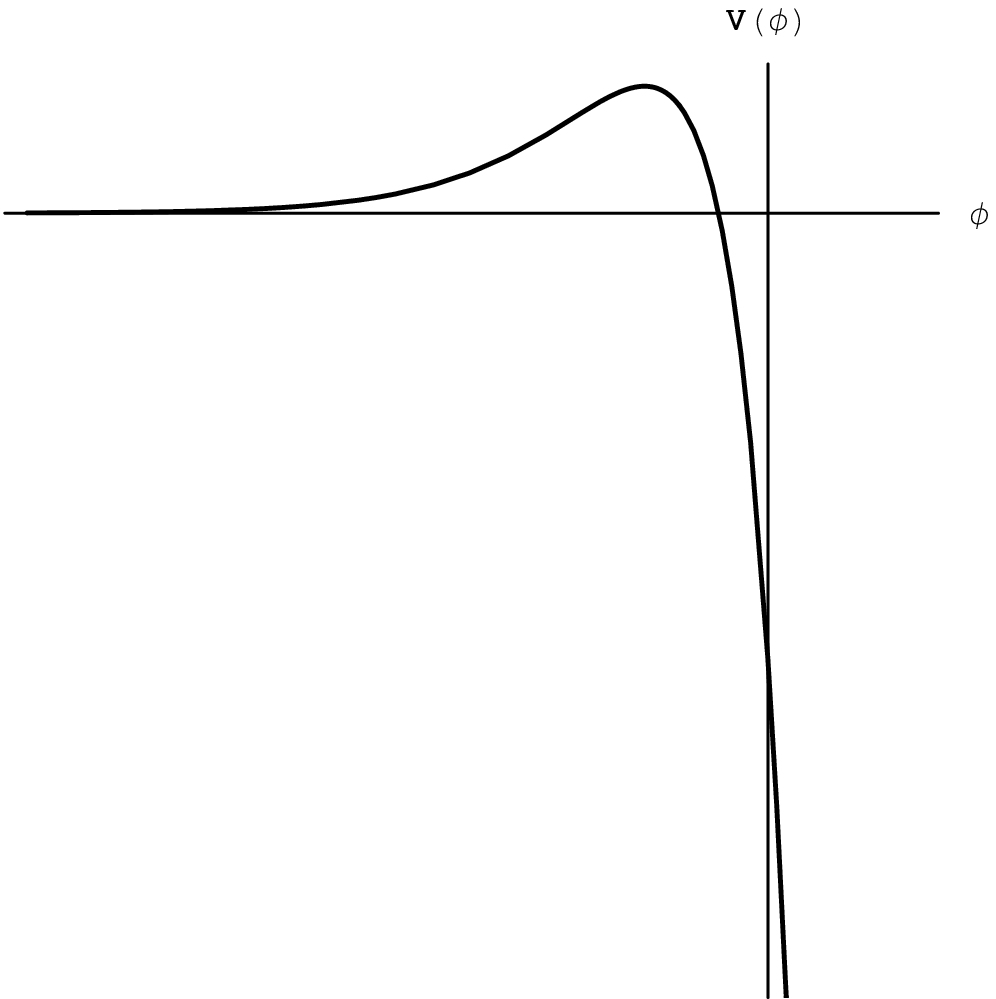}

{\small $V(\phi)$ vs $\phi$ (arbitrary units).}
\normalsize
\end{center}
Recent studies in Loop Quantum Gravity (LQG) outlined that the expectation value of the Hamiltonian operator in a given state is, in general, different from the classical Hamiltonian contribution: the application of this quantum scheme to the isotropic FRW Universe (in the presence of a massless scalar field, which plays the role of time) provided modified relations between the Hubble parameter and the energy density of the Universe. The effective cosmological dynamics is mapped into the original Friedmann equation as soon as we allow the energy density of the Universe to become negative over critical values, i.e., the following correspondence takes place
\begin{equation}\label{crit}
\epsilon\rightarrow\epsilon_{eff}\equiv \epsilon\left(1-\frac{\epsilon}{\epsilon_{crit}}\right),\ \ \epsilon_{crit}=\frac{\sqrt{3}}{16\pi^{2}\gamma^{3}}\epsilon_{pl},
\end{equation}
where $\epsilon_{crit}$ (with $\gamma$ the Immirzi parameter) is a critical value of the energy density two orders below the Planck scale, over which the matter contribution becomes negative, thus illustrating a repulsive nature of the gravitational field near the (removed) cosmological singularity. These developments can apply to the scalar-tensor model equivalent to the choice of an exponential gravitational action. In particular, as hinted by (\ref{phitronc}), a region can be found, where the potential $V(\phi)$ can be neglected. If external matter is absent, the results of LQG can apply in such a region, by modifying the Friedmann equation (\ref{frieda}), i.e.,
\begin{equation}
\left(\frac{\dot{a}}{a}\right)^{2}=\frac{8\pi G}{c^{2}}\epsilon_{eff}(\phi),
\end{equation} 
where $\epsilon_{eff}(\phi)=\epsilon(\phi)\left(1-\frac{\epsilon(\phi)}{\epsilon_{crit}}\right)$, according to (\ref{crit}). In presence of external matter, on the other hand, we obtain
\begin{equation}
\left(\frac{\dot{a}}{a}\right)^{2}=\frac{8\pi G}{c^{2}}\left(\epsilon_{eff}(\phi)+\epsilon(t)e^{-2\sqrt{\frac{16\pi G}{3c^{4}}}\phi}\right).
\end{equation}
We can now analyze the implication of the dynamical equivalence between modified gravity and scalar tensor approaches\footnote{The physical equivalence between the Jordan and the Einstein frame does not hold automatically. In fact, the transformation from one frame to the other does not assure that the physical interpretation of the solutions be the same. On the contrary, some characterizing behaviors can be lost, thus distorting the physical meaning of cosmological solutions \cite{serg}.}. In fact, the on-shell request reads $\phi=-\mu R=-\frac{R}{2\Lambda}$, and, by the power-law (\ref{scalfa}), $R=1/(3t^2)$, which does not match the solution (\ref{phitronc}) found for $\phi$, where the functional dependence on time is logarithmic. Nevertheless, as a general trend, the curvature scalar diverges as $\phi$ tends to $-\infty$,i.e., at very early times the on-shell relation is qualitatively satisfied.\\ 
This analysis shows that, near the cosmological singularity, the scalar-tensor theory takes the form of general relativity in the presence of a massless scalar field. This fact allows us to infer some hints about its quantization.
As a result, we can claim that our proposed non-Einsteinian scheme is characterized by a non-singular behavior when the corresponding scalar-tensor picture is canonically quantized. In fact, the possibility to neglect the potential field as the Big-Bang is classically approached is mapped by the results discussed in \cite{asht} into a Big-Bounce\footnote{However, a LQG formulation for the generalized $f(R)$ gravity is not yet viable and the correspondence between the Jordan and the Einstein frame on quantum level cannot be addressed. Furthermore, the non-singular feature we established here in view of the possibility to neglect the potential term near the cosmological singularity can be extended to  a wide class of  scalar-tensor theories corresponding to the $f(R)$ formulation. In particular, by the calculations above, the potential term is negligible in the asymptotic behavior towards the singularity as far as $V(\phi)$ evaluated for (\ref{phitronc}) behaves as $\mathcal{O}(t^{-2+\beta})$, with $\beta>0$. The condition on the potential term, which satisfies such a request, can be easily stated as 
\begin{equation}
\lim_{\phi\rightarrow-\infty}\frac{V(\phi)}{\phi^{\theta}e^{-2\sqrt{\frac{12\pi G}{c^{2}}}\phi}}=0,
\end{equation}
$\forall\theta>0$, while the behavior of a potential term $\sim e^{-2\phi}$ would correspond to a generalized gravitational Lagrangian linear in the $R$ variable.}. 
\subsection{$\Lambda<0$}
\begin{center}
\includegraphics[width=0.5\textwidth]{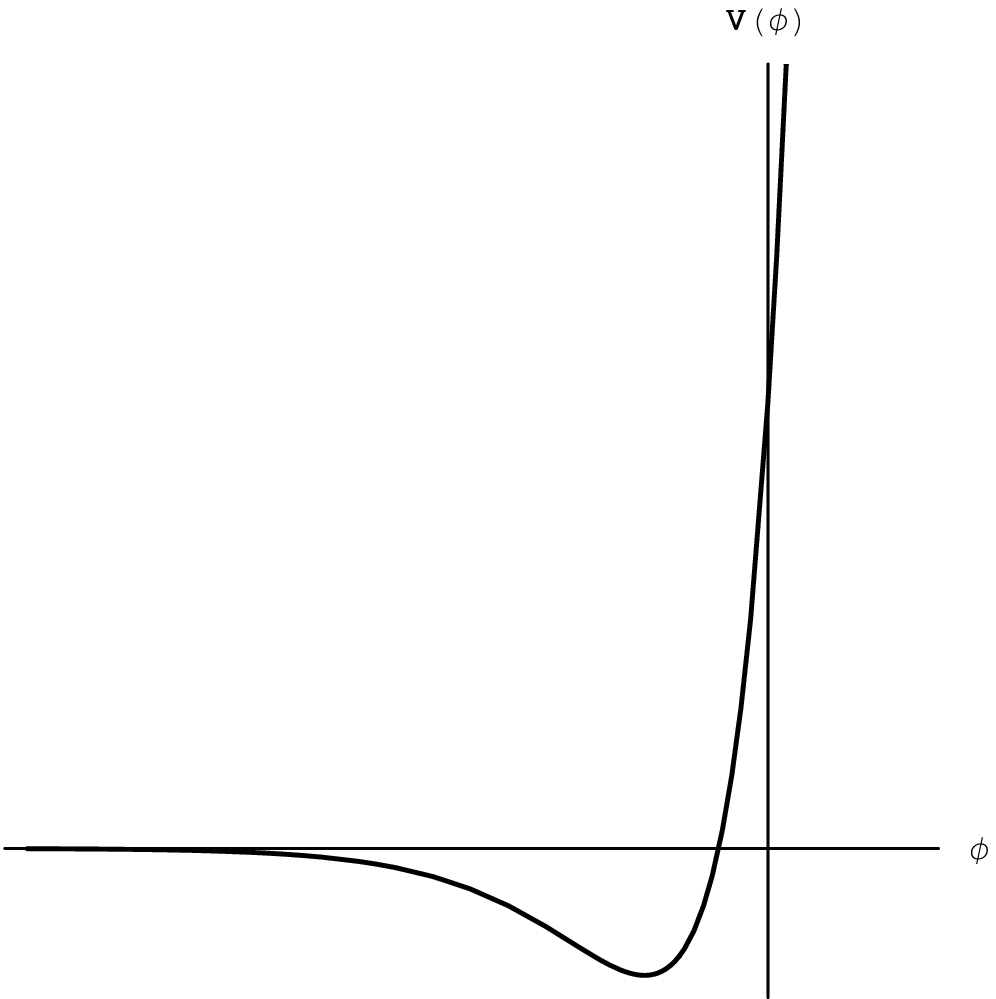}

{\small $V(\phi)$ vs $\phi$ with $\Lambda=-\mid\Lambda\mid$ (arbitrary units).}
\normalsize
\end{center}
Rel. \reff{pot} refers to a potential with no stable configuration. Reversing the sign of $\Lambda$, i.e., $\Lambda\rightarrow-\mid\Lambda\mid$, a minimum appears. This way, $\Lambda$ does not describe the cosmological term any more, but it is a parameter of the theory. For such a negative constant, the potential $V(\phi)$ admits now a minimum\footnote{The appearance of this minimum is expected to become relevant in the dynamics of the scalar field: its total energy density follows the relation
\begin{equation}
\frac{d}{dt} \left(\frac{\dot{\phi}^{2}}{2c^{2}}+V(\phi)\right)=-3H\frac{\dot{\phi}^{2}}{c^{2}}<0,
\end{equation}
where we are assuming an expanding universe, i.e., $H>0$. In fact, starting with a given value of the energy density, sooner or later, the friction due to the universe expansion settles down the scalar field near its potential minimum.} for $\phi=\phi_{min}\equiv -2\sqrt\frac{3c^{4}}{16\pi G}$, and the corresponding linearized equation reads
\begin{equation}\label{linearized}
\ddot{\phi}+3H\dot{\phi}+\frac{2}{3}c^{2}|\Lambda|e^{-2}\left(\phi+2\sqrt{\frac{3c^{4}}{16\pi G}}\right)=0.
\end{equation}
All these results can apply to the deSitter phase, so that a comparison with the previous results can be addressed. Therefore, in what follows, we search for a solution of the linearized scalar field equation (\ref{linearized}), in correspondence with the choice $a(t)=a_{0}e^{\sigma t}$ and $\epsilon(t)\equiv\varepsilon=const$.\\
The linearized equation (\ref{linearized}) then rewrites
\begin{equation}\label{linearized2}
\ddot{\phi}+3\sigma\dot{\phi}+\frac{2}{3}c^{2}|\Lambda|e^{-2}\left(\phi+2\sqrt{\frac{3c^{4}}{16\pi G}}\right)=0,
\end{equation}
whose solution around the minimum is
\begin{equation}\label{campo}
\phi=-2\sqrt{\frac{3c^{4}}{16\pi G}}+e^{-\frac{3}{2}\sigma t}\left[C_{+}\cos\beta_{+}t +C_{-}\sin\beta_{-}t\right],
\end{equation}
where $C_{\pm}$ are two arbitrary constants, and\footnote{the discriminant is negative for $x<0.24$, and, because of the prescription $x\ll1$, it is always negative, so that the field $\phi$ tends, as expected, to $\sqrt\frac{16\pi G}{3c^{4}}\phi_{min}=-2$. It is worth remarking that the on-shell relation provides the identification $\sqrt\frac{16\pi G}{3c^{4}}\phi_{min}=-\frac{6\sigma^{2}}{\Lambda c^{2}}=-2$.} $\beta_{\pm}\equiv\mp i\sqrt{|\Lambda|c^{2}}\sqrt{\frac{3x}{2}-\frac{8}{3}e^{-2}}$.\\
Solution (\ref{campo}) can now be inserted in (\ref{frieda}): since the time derivative of the scalar field can be neglected in the vicinity of the minimum, the new Friedmann equation reads
\begin{equation}\label{sigmaquadro}
\sigma^{2}=\frac{8\pi G}{3c^{2}}\left( \hat{\epsilon} e^{4}-\hat{\epsilon}_{|\Lambda|}e^{-2}\right),
\end{equation}
and can be compared with (\ref{qwert}): because of the conformal transformations (\ref{conf3}) and (\ref{conf}), the two equations completely match. Obviously, in both Jordan and Einstein frame, the metric structure remains that of a deSitter phase simply because the conformal factor $e^{-2}$ is nearly constant around the minimum\footnote{Even though the value $\phi_{min}=-2\sqrt{\frac{3c^{4}}{16\pi G}}$ would correspond to the choice $x=-R/(2\Lambda)=-2$ in the Jordan frame, such a choice can no longer be a vacuum solution of the theory in the Einstein frame. In fact, in absence of matter ($\varepsilon\equiv 0$) we would deal with a negative cosmological constant as a source of the expansion rate of the universe. However, the correspondence between the Einstein and the Jordan frame takes place as far as we compare equation (\ref{sigmaquadro}) when a constant energy density is included with relation (\ref{qwert}) obtained for $x\ll 1$. Thus we are led to postulate an off-shell correspondence between the analysis developed for a deSitter space, in which the expansion rate of the universe is much smaller than the $\mid\Lambda\mid$ value, and the scalar-tensor approach near the stable configuration, as far as matter a source is included too. The off-shell correspondence provides us with a valuable tool to regard the potential as an attractive configuration in the exponential-Lagrangian dynamics. Collecting the two points of view together, we can claim that, when dealing with an exponential Lagrangian, a deSitter phase exists, such that $\epsilon\sim\epsilon_{\mid\Lambda\mid}$ and it corresponds with general features in the space of the solution. }.
\section{cancellation}
Here we collect the issues of the previous sections together, in order to provide an explanation for the reason why the large value of the vacuum-energy density is today unobservable, or reduced to the actual cosmological constant $\mathcal{O}\left(10^{-120}\right)$ orders of magnitude smaller than it. The Jordan and the Einstein representations of the new cosmological dynamics can match only if we assume that the system evolution is always concerned with a constant matter contribution and if such a source nearly cancels the negative cosmological term, so that we fix $x\ll1$. The universal features of such a matter contribution and its constant value suggest one to identify it with the vacuum energy discussed in the previous sections. Moreover, the cancellation required to get $x\ll1$ is the natural scenario in which a relic dark energy can be recognized. By the structure of our model, the relic constant energy density must be a factor $\mathcal{O}(R/(2\mid\Lambda\mid))$ smaller than the dominant contribution $\mathcal{O}(\epsilon_{\Lambda})$. Thus if we take the vacuum energy density close to the Planckian value, then the actual ratio $R/(2\mid\Lambda\mid)$ is of order $\mathcal{O}(10^{-120})$. Such a quantity behaves like $\mathcal{O}(\frac{l_{Pl}^{2}}{L_{H}^{2}})$, where $L_{H}\sim\mathcal{O}(10^{27}cm)$ is the present Hubble radius of the universe. However, it must be remarked that such a consideration holds in the case $\epsilon_{\Lambda}$ and the vacuum energy density are the only contributions.\\  If, as below, an additional physical matter field is added, then the relic dark energy contribution is simply constrained to be less than the factor $R/(2\mid\Lambda\mid)$ of the vacuum energy. 
\subsection{Friedmann dynamics in the Einstein frame} 
Dividing the source energy density into the form
\begin{equation}
\epsilon_{mat}=\epsilon_{vac}+\rho(t),
\end{equation}
where $\rho(t)$ is a generic field contribution, then the Friedmann equation for the scalar-tensor scheme in proximity of the minimum $\phi_{min}$ reads
\begin{equation}\label{abo}
\left(\frac{\dot{a}}{a}\right)^{2}=\frac{8\pi G}{3c^{2}}\left( (\hat{\epsilon}_{vac}+\hat{\rho}(t) )e^{4}-\hat{\epsilon}_{|\Lambda|}e^{-2}\right).
\end{equation}
Since the compatibility of the Jordan- and the Einstein-frame approaches requires that the expansion rate of the Universe be much smaller than the corresponding parameter $\Lambda$, then we are led to account for the non-exact cancellation of the vacuum-energy density by the small parameter $\delta\ll1$ as follows.\\
If we take $\hat{\epsilon}_{vac}e^{4}=e^{-2}\hat{\epsilon}_{\mid\Lambda\mid}(1+\frac{\delta}{2})=\pi e^{4}\alpha^{4}\epsilon_{pl}$, i.e., $\mid\Lambda\mid\sim 8\pi^{2}\alpha^{4} e^{6}/l^{2}_{pl}$, eq. (\ref{abo}) restates
\begin{equation}\label{albero}
\left(\frac{\dot{a}}{a}\right)^{2}=\frac{8\pi G}{3c^{2}}\left( \hat{\rho}(t)e^{4}-\hat{\epsilon}_{|\Lambda|}e^{-2}\frac{\delta}{2}\right).
\end{equation}
Thus, when the constant energy density dominates, we recognize $\delta=\mathcal{O}\left(\frac{l_{Pl}^{2}}{L_{H}^{2}}\right)$, since now $\epsilon_{\mid\Lambda\mid}$ has a Planckian value. We note that the factor $e^{6}$ appearing in the expression of $\Lambda$ is of course present only in the scalar-tensor theory, because of the rescaling of the involved energy densities.
\subsection{Friedmann dynamics in the Jordan frame}
On the other hand, this picture can be recovered even in the original Jordan frame, as far as we observe that, for a Planckian value of $\Lambda$, the exponential Lagrangian is expandable in power series immediately after the Planckian era of the Universe. In fact, as far as we fix $\epsilon_{\Lambda}$ at Planckian scales, then, as emphasized above, we automatically get for $\delta\equiv x$ of order $\mathcal{O}(10^{-120})$. If we now introduce a pure matter contribution, $\epsilon_{mat}\ll \mid \epsilon_{vac} \mid$, it is easy to recognize that the standard Friedmann equation with the present cosmological constant is recovered:
\begin{equation}
H_{0}^{2}=\frac{\Lambda c^{2}}{6}\left(\frac{\epsilon_{vac}+\epsilon_{mat}}{\epsilon_{\Lambda}}+1\right)= \frac{8\pi G}{3c^{2}}\epsilon_{mat}+\frac{\delta\Lambda c^{2}}{6}.
\end{equation}
All our considerations refer here to the deSitter solution, and, therefore, $\epsilon_{mat}$ is to be regarded as constant. However, it is naturally expected that the Friedmann equation with a small cosmological term arises as low-energy curvature of this theory for any dependence on $\epsilon_{mat}$; in fact, for our choice of $\epsilon_{\Lambda}$, the Lagrangian density of the gravitational field explicitly reads
\begin{equation}
L=\frac{\hbar}{l_{pl}^{4}}\pi\alpha^{4}e^{\frac{-Rl_{pl}^{2}}{16\pi^{2}\alpha^{4}}}.
\end{equation}
From this expression for the gravitational-field Lagrangian density, we recognize that, as far as the typical length scale $\mathcal{D}\gg l_{pl}$ of the curvature ($R\sim1/\mathcal{D}^{2}$), we can address the expansion in terms of small quantity $l_{pl}^{2}/\mathcal{D}^{2}$
\begin{equation}
L\simeq \frac{\hbar}{l_{pl}^{4}}\pi\alpha^{4}-\frac{\hbar}{16\pi l_{pl}}R+\mathcal{O}\left(\frac{1}{\mathcal{D}^{2}}\right)
\end{equation}
This approximated Lagrangian density would provide for the FRW metric the following Friedmann equation\footnote{Such an approximated equation, isomorphic to (\ref{albero}), allows us to reproduce all the considerations developed about the exact deSitter case. However, the analysis performed above is relevant in the Jordan frame in outlining the necessity of the constraint $\epsilon_{vac}/\epsilon_{\Lambda}\sim-1$.\\
In fact, the exact deSitter case clarified that, for positive $\Lambda$ values, this relation is the only one able to provide the consistency of the Friedmann equation according to an exponential Lagrangian density. This feature could not be recognized by an approximated analysis, as in (\ref{free}).}
\begin{equation}\label{free}
\left(\frac{\dot{a}}{a}\right)^{2}=\frac{8\pi G}{3c^{2}}\left[\epsilon_{mat}(t)+\epsilon_{\Lambda}\left(\frac{\epsilon_{vac}}{\epsilon_{\Lambda}}+1\right) \right].
\end{equation}
\subsection{Discussion}
After deriving the Einstein equations for a generalized gravitational action and specifying the results for an FRW metric, the particular choice of an exponential Lagrangian density has been analyzed.\\
The free parameters of such a Lagrangian density have been fixed as functions of the cosmological constant, and, in the deSitter regime, the ratio between the vacuum-energy density and the geometrical contribution has been illustrated to acquire a negative sign, which has been the springboard for the investigation of the relation between the vacuum-energy and cut-off approaches to the geometrical description of the Universe. In particular, the cut-off introduced in the vacuum-energy density has been linked with the modified commutation relation following from a generalized uncertainty principle, and has been fixed at Planck scales.\\
The negative sign of the ratio $\epsilon_{vac}/\epsilon_{\Lambda}$ not only explains the non-observability of the cut-off vacuum-energy density and is in line with the LQC prediction of the Big Bounce in an FRW metric, but also allows one to recover the standard Friedmann equation in the deSitter phase, when the matter contribution is taken into account, and for any choice of the matter terms.\\
Studying some aspects of the pertinent scalar-tensor description has allowed us to investigate further connotations of the implementation of such a scheme. In particular, the physical meaning of the sign of the cosmological constant has been explained to provide interesting hints about cosmological implications.
The main issue of our analysis has consisted in fixing the link between the vacuum-energy cancellation and the present Universe Dark Energy: the actual acceleration, observed via SNIA, is due to the relic of the original huge vacuum energy, after its mean value has been compensated for by the intrinsic cosmological constant $\Lambda$ contained in the exponential Lagrangian.

\end{document}